\documentclass[12pt]{iopart}
\usepackage{graphicx}
\usepackage[font=small,format=plain,labelfont=bf,up,textfont=normal,up]{caption}
\usepackage{dblfloatfix}
\usepackage{titlesec}

\titleformat{\subsection}[block]{\hspace{2em}}{\flushright\textbf\thesubsection}{1em}{\textbf}
\begin{document}

\title{SppC based energy frontier lepton-proton colliders: luminosity and physics}

\author{Ali Can Canbay$^{1,2}$, Umit Kaya$^{1, 2, *}$, Bora Ketenoglu$^{3}$, Bilgehan Baris Oner$^{1}$, Saleh Sultansoy$^{1,4}$}

\address{$^{1}$TOBB University of Economics and Technology, Ankara, Turkey \\
$^{2}$Ankara University, Department of Physics, Ankara, Turkey \\
$^{3}$Ankara University, Department of Engineering Physics, Ankara, Turkey \\
$^{4}$ANAS Institute of Physics, Baku, Azerbaijan}

\vspace{10pt}
\begin{indented}
\item[]$^{*}$Correspondence: umit.kaya@cern.ch
\end{indented}
\vspace{10pt}
\begin{indented}
\item[]April 2017
\end{indented}

\begin{abstract}
In this study, main parameters of Super proton-proton Collider (SppC) based lepton-proton colliders are estimated. For electron beam parameters, highest energy International Linear Collider (ILC) and Plasma Wake Field Accelerator-Linear Collider (PWFA-LC) options are taken into account. For muon beams, 1.5 TeV and 3 TeV center of mass energy Muon Collider parameters are used. In addition, ultimate $\mu$p collider which assumes construction of additional 50 TeV muon ring in the SppC tunnel is considered as well. It is shown that luminosity values exceeding $10^{32}$ $cm^{-2}s^{-1}$ can be achieved with moderate upgrade of the SppC proton beam parameters. Physics search potential of proposed lepton-proton colliders is illustrated by considering small Bj{\"o}rken x region as an example of SM physics and resonant production of color octet leptons as an example of BSM physics.              
\end{abstract}

\providecommand{\keywords}[1]{\textbf{\textit{\small Keywords:}} #1}

\section{Introduction}

It is known that lepton-hadron scattering had played crucial role
in our understanding of deep inside of matter. For example, electron
scattering on atomic nuclei reveals structure of nucleons in Hofstadter
experiment [1]. Moreover, quark parton model was originated
from lepton-hadron collisions at SLAC [2]. Extending the
kinematic region by two orders of magnitude both in high $Q^{2}$
and small $x$, HERA (the first and still unique lepton-hadron collider)
with $\sqrt{s}=0.32$ TeV has shown its superiority compared to the
fixed target experiments and provided parton distribution functions
(PDF) for LHC and Tevatron experiments (for review of HERA results see [3, 4]). Unfortunately, the region
of sufficiently small $x$ ($<10^{-5}$) and high $Q^{2}$ ($\geq10\,GeV^{2}$) simultaneously,
where saturation of parton densities should manifest itself, has not
been reached yet. Hopefully, LHeC [5] with $\sqrt{s}=1.3$
TeV will give opportunity to touch this region. 

Construction of linear $e^{+}e^{-}$colliders (or dedicated linac) and
muon colliders (or dedicated muon ring) tangential to the future circular pp
colliders, FCC or SppC, as shown in Fig. 1, will give opportunity to use
highest energy proton beams in order to obtain highest center of mass energy in lepton-hadron and photon-hadron collisions. (For earlier studies on linac-ring type ep, $\gamma$p, eA and $\gamma$A colliders, see reviews [6, 7] and papers [8-14].)

\begin{figure}[hbtp]
\centering
\includegraphics[scale=0.45]{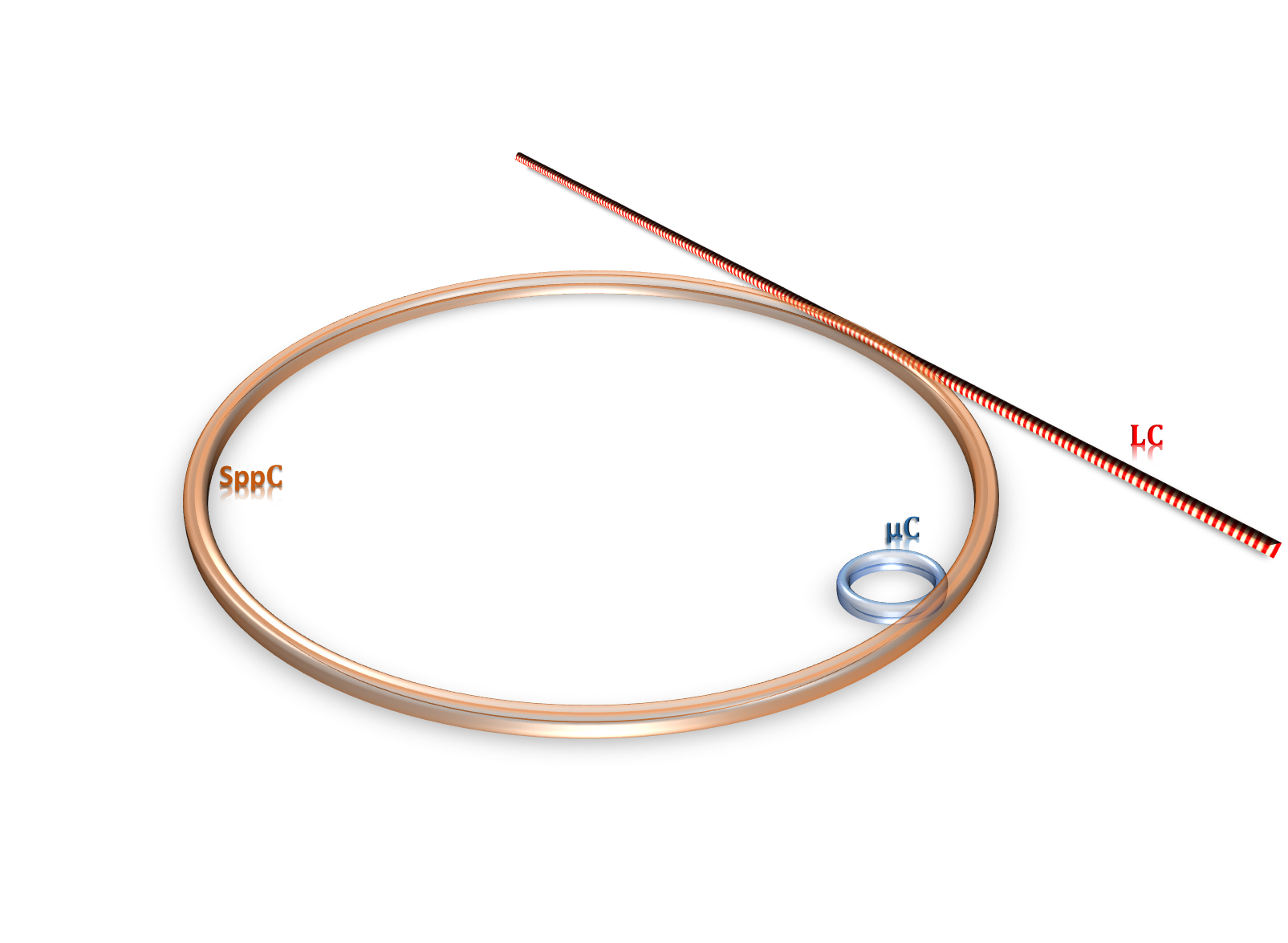}
\caption{Possible configuration for SppC, linear collider (LC) and muon collider (${\mu}C)$.}
\end{figure}

FCC is the future 100 TeV center-of-mass energy pp collider studied at CERN and supported by European Union within the Horizon 2020 Framework Programme for Research and Innovation [15]. SppC is the Chinese analog of the FCC. Main parameters of the SppC proton beam [16, 17] are presented in  Table~\ref{tab:tablo1}. The FCC based ep and $\mu$p colliders have been considered recently (see [18] and references therein).
\clearpage

\begin{table}[!h]
\captionsetup{singlelinecheck=false, justification=justified}      
\caption{Main parameters of proton beams in SppC.}\label{tab:tablo1}
\centering
\begin{tabular}{|c|c|c|}
\hline 
Beam Energy (TeV) & 35.6 & 68.0 \\ 
\hline 
Circumference (km) & 54.7 & 100.0 \\
\hline 
Peak Luminosity ($10^{34}\,cm^{-2}s^{-1}$)  & 11 & 102 \\ 
\hline 
Particle per Bunch ($10^{10}$) & 20 & 20 \\ 
\hline 
Norm. Transverse Emittance ($\mu m$) & 4.10 & 3.05 \\ 
\hline 
{$\beta$}{*} amplitude function at IP (m)  & 0.75 & 0.24 \\ 
\hline 
IP beam size ($\mu m$) & 9.0 & 3.04 \\ 
\hline 
Bunches per Beam  & 5835 & 10667 \\ 
\hline 
Bunch Spacing (ns) & 25 & 25 \\ 
\hline 
Bunch length (mm) & 75.5 & 15.8 \\ 
\hline 
Beam-beam parameter, $\xi_{pp}$  & 0.006 & 0.008 \\ 
\hline 
\end{tabular}
\end{table}
\vspace{20pt}
   
In this paper we consider SppC based ep and $\mu$p colliders. In Section 2, main parameters of proposed colliders, namely center of mass energy and luminosity, are estimated taken into account beam-beam tune shift and disruption effects. Physics search potential of the SppC based lp colliders have been evaluated in Section 3, where small Bj{\"o}rken-x region is considered as an example of the SM physics and resonant production of color octet leptons is considered as an example of the BSM physics. Our conclusions and recommendations are presented in Section 4.\\

\section{Main Parameters of the SppC Based ep and $\mu$p Colliders}

\vspace{10pt}

General expression for luminosity of SppC based $lp$ colliders is
given by ($l$ denotes electron or muon):

\begin{eqnarray}
L_{lp} & = & \frac{N_{l}N_{p}}{4\pi max[\sigma_{x_{p}},\sigma_{x_{l}}]max[\sigma_{y_{p}},\sigma_{y_{l}}]}min[f_{c_{p},}\,f_{c_{l}}]\label{eq:Denklem1}
\end{eqnarray}\\

\noindent where $N_{l}$ and $N_{p}$ are numbers of leptons and protons per
bunch, respectively; $\sigma_{x_{p}}$ ($\sigma_{x_{l}}$) and $\sigma_{y_{p}}$
($\sigma_{y_{l}}$) are the horizontal and vertical proton (lepton)
beam sizes at interaction point (IP); $f_{c_{l}}$ and $f_{c_{p}}$ are LC/$\mu$C and SppC bunch
frequencies. $f_{c}$ is expressed by $f_{c}=N_{b}f_{rep}$, where
$N_{b}$ denotes number of bunches, $f_{rep}$ means revolution frequency
for SppC/$\mu$C and pulse frequency for LC. In order to determine collision
frequency of lp collider, minimum value should be chosen among lepton
and hadron bunch frequencies. Some of these parameters can be rearranged
in order to maximize $L_{lp}$ but one should note that there are
main limitations due to beam-beam effects that should be kept in mind. While
beam-beam tune shift affects proton and muon beams,
disruption has influence on electron beams.

Disruption parameter for electron beam is given by: 

\begin{eqnarray}
D_{x_{e}} & = & \frac{2\,N_{p}r_{e}\sigma_{z_{p}}}{\gamma_{e}\sigma_{x_{p}}(\sigma_{x_{p}}+\sigma_{y_{p}})}\label{eq:Denklem2}
\end{eqnarray}

$\,$

\begin{equation}
D_{y_{e}}=\frac{2\,N_{p}r_{e}\sigma_{z_{p}}}{\gamma_{e}\sigma_{y_{p}}(\sigma_{y_{p}}+\sigma_{x_{p}})}
\end{equation}

\vspace{10pt}

\noindent where, $r_{e}=2.82\times10^{-15}$ $m$ is classical radius for
electron, $\gamma_{e}$ is the Lorentz factor of electron beam, $\sigma_{x_{p}}$
and $\sigma_{y_{p}}$ are horizontal and vertical proton beam sizes
at IP, respectively. $\sigma_{z_{p}}$ is bunch length of proton beam. Beam-beam parameter for proton beam is given by:\\

\begin{equation}
\xi_{x_{p}}=\frac{N_{l}r_{p}\beta_{p}^{*}}{2\pi\gamma_{p}\sigma_{x_{l}}(\sigma_{x_{l}}+\sigma_{y_{l}})}\label{eq:Denklem3}
\end{equation}

$ $

\begin{equation}
\xi_{y_{p}}=\frac{N_{l}r_{p}\beta_{p}^{*}}{2\pi\gamma_{p}\sigma_{y_{l}}(\sigma_{y_{l}}+\sigma_{x_{l}})}
\end{equation}

\vspace{10pt}

\noindent where $r_{p}$ is classical radius for proton,
$r_{p}=1.54\times10^{-18}$ $m$, $\beta_{p}^{*}$ is beta function of
proton beam at IP, $\gamma_{p}$ is the Lorentz
factor of proton beam. $\sigma_{x_{l}}$ and $\sigma_{y_{l}}$ are
horizontal and vertical sizes of lepton beam at IP, respectively.\\

Beam-beam parameter for muon beam is given by:\\

\begin{equation}
\xi_{x_{\mu}}=\frac{N_{p}r_{\mu}\beta_{\mu}^{*}}{2\pi\gamma_{\mu}\sigma_{x_{p}}(\sigma_{x_{p}}+\sigma_{y_{p}})}\label{eq:Denklem3}
\end{equation}

$ $

\begin{equation}
\xi_{y_{\mu}}=\frac{N_{p}r_{\mu}\beta_{\mu}^{*}}{2\pi\gamma_{\mu}\sigma_{y_{p}}(\sigma_{y_{p}}+\sigma_{x_{p}})}
\end{equation}

\vspace{10pt}

\noindent where $r_{\mu}=1.37\times10^{-17}$ $m$ is classical muon radius, $\beta_{\mu}^{*}$ is beta function of
muon beam at IP, $\gamma_{\mu}$ is the Lorentz
factor of muon beam. $\sigma_{x_{p}}$ and $\sigma_{y_{p}}$ are
horizontal and vertical sizes of proton beam at IP, respectively.

\subsection{ep option}

Preliminary study of CepC-SppC based e-p collider with $\sqrt{s}=4.1$ TeV and $L_{ep}=10^{33}$ $cm^{-2}s^{-1}$ has been performed in [19]. In this subsection, we consider ILC (International Linear Collider) [20] and PWFA-LC (Plasma Wake Field Accelerator - Linear Collider) [21] as a source of electron/positron beam for SppC based energy frontier ep colliders. Main parameters of ILC and PWFA-LC electron beams are given Table~\ref{tab:tablo2}.
\clearpage

\begin{table}[!h]
\captionsetup{singlelinecheck=false, justification=justified}  
\caption{ Main parameters of the ILC (second column) and PWFA-LC (third column) electron beams.}\label{tab:tablo2}
\centering
\begin{tabular}{|c|c|c|}
\hline 
Beam Energy (GeV)  & $500$  & $5000$\tabularnewline
\hline 
Peak Luminosity ($10^{34}\,cm^{-2}s^{-1}$)  & $4.90$  & $6.27$\tabularnewline
\hline 
Particle per Bunch ($10^{10}$)  & $1.74$  & $1.00$\tabularnewline
\hline 
Norm. Horiz. Emittance ($\mu m$)  & $10.0$  & $10.0$\tabularnewline
\hline 
Norm. Vert. Emittance (nm)  & $30.0$  & $35.0$\tabularnewline
\hline 
Horiz. {$\beta$}{*}  amplitude function at IP (mm)  & $11.0$  & $11.0$\tabularnewline
\hline 
Vert. {$\beta$}{*}  amplitude function at IP (mm)  & $0.23$  & $0.099$\tabularnewline
\hline 
Horiz. IP beam size (nm)  & $335$  & $106$\tabularnewline
\hline 
Vert. IP beam size (nm)  & $2.70$  & $59.8$\tabularnewline
\hline 
Bunches per Beam  & $2450$  & $1$\tabularnewline
\hline 
Repetition Rate (Hz)  & $4.00$  & $5000$\tabularnewline
\hline 
Beam Power at IP (MW)  & $27.2$  & $40$\tabularnewline
\hline 
Bunch Spacing (ns)  & $366$  & $20$x$10^{4} $\tabularnewline
\hline 
Bunch length (mm)  & $0.225$  & $0.02$\tabularnewline
\hline 
\end{tabular}

\end{table}
\vspace{20pt}

It is seen that bunch spacings of ILC and PWFA-LC are much greater than SppC bunch spacing. On the other hand, transverse size of proton beam is much greater than transverse sizes of electron beam. Therefore, Eq. (1) for luminosity turns into:

\begin{equation}
L_{ep}=\frac{N_{e}N_{p}}{4\pi\sigma_{p}^{2}}f_{c_{e}}\label{eq:Denklem4}
\end{equation}

\vspace{10pt}

For transversely matched electron and proton beams at IP, equations for electron beam disruption and proton beam tune shift become:

\begin{equation}
D_{e}=\frac{N_{p}r_{e}\sigma_{z_{p}}}{\gamma_{e}\sigma_{p}^{2}}\label{eq:Denklem5}
\end{equation}

\begin{equation}
\xi_{p}=\frac{N_{e}r_{p}\beta_{p}^{*}}{4\pi\gamma_{p}\sigma_{p}^{2}}=\frac{N_{e}r_{p}}{4\pi\epsilon_{np}}\label{eq:Denklem6}
\end{equation}

\noindent where $\epsilon_{np}$ is normalized transverse emittance of proton beam.

Using nominal parameters of ILC, PWFA-LC and SppC, we obtain values of L$_{ep}$, D$_e$ and ${\xi}_{p}$ parameters for LC$\otimes$SppC based ep colliders, which are given in Table~\ref{tab:tablo3}. The values for luminosity given in parantheses represent results of beam-beam simulations by ALOHEP software [22], which is being developed for linac-ring type ep colliders.
\clearpage

\begin{table}[!h]
\captionsetup{singlelinecheck=false, justification=justified}  
\caption{  Main parameters of LC$\otimes$SppC based ep colliders.}\label{tab:tablo3}
\centering
\begin{tabular}{|c|c|c|c|c|c|}
\hline 
E$_{e}$, TeV & E$_{p}$, TeV & $\sqrt{s}$, TeV & L$_{ep}$, $cm^{-2}s^{-1}$ & D$_{e}$ & $\xi_{p}$, $10^{-3}$ \\ 
\hline 
0.5 & 35.6 & 8.44 & 3.35 (6.64) x $10^{30}$ & 0.537 & 0.5 \\ 
\hline 
0.5 & 68 & 11.66 & 2.69 (5.33) x $10^{31}$ & 0.902 & 0.7 \\ 
\hline 
5 & 35.6 & 26.68 & 0.98 (1.94) x $10^{30}$ & 0.054 & 0.3 \\ 
\hline 
5 & 68 & 36.88 & 0.78 (1.56) x $10^{31}$ & 0.090 & 0.4 \\ 
\hline 
\end{tabular} 
\end{table} 
\vspace{10pt}

In order to increase luminosity of ep collisions LHeC-like upgrade of the SppC proton beam parameters have been used. Namely, $\beta$ function of proton beam at IP is arranged to be 7.5/2.4 times lower (0.1 m instead of 0.75/0.24 m) which corresponds to LHeC [5] and THERA [23] designs. This leads to increase of luminosity and D$_{e}$ by factor 7.5 and 2.4 for SppC with 35.6 TeV and 68 TeV proton beam, respectively. Results are shown in Table~\ref{tab:tablo4}.  

\begin{table}[!h]
\captionsetup{singlelinecheck=false, justification=justified}  
\caption{  Main parameters of LC$\otimes$SppC based ep colliders with upgraded $\beta$*.}\label{tab:tablo4}
\centering
\begin{tabular}{|c|c|c|c|c|c|}
\hline 
E$_{e}$, TeV & E$_{p}$, TeV & $\sqrt{s}$, TeV & L$_{ep}$, $cm^{-2}s^{-1}$ & D$_{e}$ & $\xi_{p}$, $10^{-3}$ \\ 
\hline 
0.5 & 35.6 & 8.44 & 2.51 (4.41) x $10^{31}$ & 4.03 & 0.5 \\ 
\hline 
0.5 & 68 & 11.66 & 6.45 (10.8) x $10^{31}$ & 2.16 & 0.7 \\ 
\hline 
5 & 35.6 & 26.68 & 7.37 (13.3) x $10^{30}$ & 0.403 & 0.3 \\ 
\hline 
5 & 68 & 36.88 & 1.89 (3.75) x $10^{31}$ & 0.216 & 0.4 \\ 
\hline 
\end{tabular} 
\end{table} 
\vspace{10pt}

In principle "dynamic focusing scheme" [24] which was proposed for THERA, could provide additional factor of 3-4. Therefore, luminosity values exceeding $10^{32}$ $cm^{-2}s^{-1} $ can be achieved for all options.
Concerning ILC$\otimes$SppC based ep colliders, a new scheme for energy recovery proposed for higher-energy LHeC (see Section 7.1.5 in [5]) may give an opportunity to increase luminosity by an additional order, resulting in L$_{ep}$ exceeding $10^{33}$ $cm^{-2}s^{-1} $. Unfortunately, this scheme can not be applied at PWFA-LC$\otimes $SppC.

\subsection{$\mu$p option}

Muon-proton colliders were proposed almost two decades ago: construction of additional proton ring in $\sqrt{s}$ = 4 TeV muon collider tunnel was suggested in [25], construction of additional 200 GeV energy muon ring in the Tevatron tunnel was  considered in [26] and ultimate $\mu$p collider with 50 TeV proton ring in $\sqrt{s}$ = 100 TeV muon collider tunnel was suggested in [27]. Here, we consider construction of TeV energy muon colliders ($\mu$C) [28]  tangential to the SppC. Parameters of $\mu$C are given in Table~\ref{tab:tablo5}. 

Keeping in mind that both SppC and $\mu$C have round beams, luminosity Eq. (1) turns to:

\begin{eqnarray}
L_{pp} & = & f_{pp}\frac{N_{p}^{2}}{4\pi\sigma_{p}^{2}}\label{eq:Denklem7}
\end{eqnarray}

\begin{center}
\begin{eqnarray}
L_{\mu\mu} & = & f_{\mu\mu}\frac{N_{\mu}^{2}}{4\pi\sigma_{\mu}^{2}}\label{eq:Denklem8}
\end{eqnarray}
 
\par\end{center}

\noindent for SppC-$pp$ and $\mu$C, respectively. Concerning muon-proton
collisions one should use larger transverse beam sizes and smaller
collision frequency values. Keeping in mind that $f_{\mu\mu}$ is
smaller than $f_{pp}$ by more than two orders, following correlation between $\mu p$
and $\mu\mu$ luminosities take place:
\begin{center}
\begin{eqnarray}
L_{\mu p} & = & (\frac{N_{p}}{N_{\mu}})(\frac{\sigma_{\mu}}{max[\sigma_{p},\,\sigma_{\mu}]})^{2}L_{\mu\mu}\label{eq:Denklem9}
\end{eqnarray}
\par\end{center}

Using nominal parameters of $\mu\mu$ colliders given in Table 5,
parameters of the SppC based
$\mu p$ colliders are calculated according to Eq. (\ref{eq:Denklem9}) and presented in Table~\ref{tab:tablo6}. Concerning beam beam tune shifts, for round and matched beams Eqs. (4,5) and Eqs. (6,7) turns to:

\begin{eqnarray}
\xi_{p}  =  \frac{N_{\mu}r_{p}\beta_{p}^{*}}{4\pi\gamma_{p}\sigma_{\mu}^{2}}  = \frac{N_{\mu}r_{p}}{4\pi\epsilon_{np}}\label{eq:Denklem10}
\end{eqnarray}

\noindent and

\begin{eqnarray}
\xi_{\mu}  =  \frac{N_{p}r_{\mu}\beta_{\mu}^{*}}{4\pi\gamma_{\mu}\sigma_{p}^{2}}  = \frac{N_{p}r_{\mu}}{4\pi\epsilon_{n\mu}},\label{eq:Denklem11}
\end{eqnarray}

\noindent respectively.

As one can see from Table~\ref{tab:tablo6}, where nominal parameters of SppC proton beam are used, $\xi_{p}$ is unacceptably high and should be decreased to 0.02 which seems acceptable for $\mu$p colliders [26]. According to Eq. (14), $\xi_{p}$ can be decreased, for example, by decrement of N$_{\mu}$ which leads to corresponding reduction of luminosity (three times and four times for $\mu$p 35.6 TeV and 68 TeV, respectively). Alternatively, crab crossing [29] can be used for decreasing of $\xi_{p}$ without change of the luminosity.

\begin{table}[!h]
\captionsetup{singlelinecheck=false, justification=justified}  
\caption{Main parameters of the muon beams.}\label{tab:tablo5}
\centering
 
\begin{tabular}{|c|c|c|}
\hline 
Beam Energy (GeV)  & $750$  & $1500$  \tabularnewline
\hline 
Circumference (km)  & $2.5$  & $4.5$\tabularnewline
\hline  
Average Luminosity ($10^{34}\,cm^{-2}s^{-1}$)  & $1.25$  & $4.4$\tabularnewline
\hline 
Particle per Bunch ($10^{12}$)  & $2$  & $2$\tabularnewline
\hline 
Norm. Trans. Emitt. (mm-rad)  & $0.025$  & $0.025$\tabularnewline
\hline  
{$\beta$}{*}  amplitude function at IP (cm)  & $1 (0.5-2)$  & $0.5 (0.3-3)$\tabularnewline
\hline 
IP beam size ($\mu$m)  & $6$  & $3$\tabularnewline
\hline 
Bunches per Beam  & $1$  & $1$\tabularnewline
\hline 
Repetition Rate (Hz)  & $15$  & $12$\tabularnewline
\hline
Bunch Spacing (ns)  & $8300$  & $15000$\tabularnewline
\hline 
Bunch length (cm)  & $1$  & $0.5$\tabularnewline
\hline 
\end{tabular}
\end{table}
 
\begin{table}[!h]
\captionsetup{singlelinecheck=false, justification=justified}  
\caption{Main parameters of SppC based $\mu$p colliders.}\label{tab:tablo6}
\centering     

\begin{tabular}{|c|c|c|c|c|c|}
\hline 
$E_{\mu}$, TeV & $E_{p}$, TeV & $\surd$S, TeV & $L_{{\mu}p} $, $cm^{-2}s^{-1}$ & $\xi_{\mu}$ & $\xi_{p}$ \\ 
\hline 
0.75 & 35.6 & 10.33 & 5.5 x $10^{32}$ &  8.7 x $10^{-3}$ &  6.0 x $10^{-2}$ \\ 
\hline 
0.75 & 68 & 14.28 & 12.5 x $10^{32}$ &  8.7 x $10^{-3}$ &  8.0 x $10^{-2}$ \\ 
\hline 
1.5 & 35.6 & 14.61 & 4.9 x $10^{32}$ &  8.7 x $10^{-3}$ &  6.0 x $10^{-2}$ \\ 
\hline 
1.5 & 68 & 20.2 & 42.8 x $10^{32}$ &  8.7 x $10^{-3}$ &  8.0 x $10^{-2}$ \\ 
\hline 
\end{tabular} 
\end{table}
  
\subsection{Ultimate $\mu$p option}

This option can be realized if an additional muon ring is constructed in the SppC tunnel. In order to estimate CM energy and luminosity of $\mu$p collisions we use muon beam parameters from [30], where 100 TeV center of mass energy muon collider with 100 km ring circumference have been proposed. These parameters are presented in Table~\ref{tab:tablo7}.

\vspace{10pt}

CM energy, luminosity and tune shifts for ultimate $\mu$p collider are given in Table~\ref{tab:tablo8}. Again $\xi_{\mu}$ and $\xi_{p}$ can be decreased by lowering of $N_{p}$ and $N_{\mu}$ respectively (which lead to corresponding decrease of luminosity) or crab crossing can be used without change of the luminosity. 

\begin{table}[!ht]
\captionsetup{singlelinecheck=false, justification=justified}  
\caption{Main parameters of the ultimate muon beam.}\label{tab:tablo7}
\centering  
\begin{tabular}{|c|c|}
\hline 
Beam Energy (TeV)  & $50$    \tabularnewline
\hline 
Circumference (km)  & $100$ \tabularnewline
\hline  
Average Luminosity ($10^{34}\,cm^{-2}s^{-1}$)  & $100$ \tabularnewline
\hline 
Particle per Bunch ($10^{12}$)  & $0.80$ \tabularnewline
\hline 
Norm. Trans. Emitt. (mm-mrad)  & $8.7$  \tabularnewline
\hline  
{$\beta$}{*}  amplitude function at IP (mm)  & $2.5$ \tabularnewline
\hline 
IP beam size ($\mu$m)  & $0.21$ \tabularnewline
\hline 
Bunches per Beam  & $1$  \tabularnewline
\hline 
Repetition Rate (Hz)  & $7.9$  \tabularnewline
\hline
Bunch Spacing ($\mu$s)  & $333$  \tabularnewline
\hline 
Bunch length (mm)  & $2.5$  \tabularnewline
\hline 
\end{tabular}
\end{table}

\begin{table}[!ht]
\captionsetup{singlelinecheck=false, justification=justified}  
\caption{Main parameters of the ultimate SppC based ${\mu}$p collider.}\label{tab:tablo8}
\centering 

\begin{tabular}{|c|c|c|c|c|c|}
\hline  
$E_{\mu}$, TeV & $E_{p}$, TeV & $\surd$S, TeV & $L_{{\mu}p} $, $cm^{-2}s^{-1}$ & $\xi_{\mu}$ & $\xi_{p}$ \\ 
\hline 
50 & 68 & 116.6 & 1.2 x $10^{33}$ &  2.6 x $10^{-2}$ &  3.5 x $10^{-2}$ \\ 
\hline  
\end{tabular} 
\end{table} 

\section{Physics}

In order to evaluate physics search potential of the SppC based lp colliders we consider two phenomena, namely, small Bj{\"o}rken $x$ region is considered as an example of the SM physics and resonant production of color octet electron and muon is considered as an example of the BSM physics.

\subsection{Small Bj{\"o}rken $x$}

As mentioned above, investigation of extremely small $x$ region ($x$ $<$ $10^{-5}$) at sufficiently large $Q^{2}$ ($>$ 10 $GeV^{2}$), where saturation of parton density should manifest itself, is crucial for understanding of QCD basics. Smallest achievable $x$ at lp colliders is given by $Q^{2}$/S. For LHeC with $\sqrt{s}=1.3$ TeV minimal acvievable value is $x$ = 6 x $10^{-6}$. In Table~\ref{tab:tablo9}, we present smallest $x$ values for different SppC based lepton-proton colliders (E$_{p}$ is chosen as 68 TeV). It is seen that proposed machines has great potential for enligthening of QCD basics. 

\begin{table}[!h]
\captionsetup{singlelinecheck=false, justification=justified}  
\caption{Attainable Bj{\"o}rken $x$ values at $Q^{2}=10$ $GeV^{2}$.}\label{tab:tablo9}
\centering 

\begin{tabular}{|c|c|c|c|c|}
\hline 
 E$_{l}$ (TeV) & 0.5   & 5 & 1.5  & 50 \\ 
\hline 
 $x$ & $7$ x $10^{-8}$  & $7$ x $10^{-9}$  &$2$ x $10^{-8}$   &  $7$ x $10^{-10}$  \\ 
  \hline 
\end{tabular} 
\end{table}

\subsection{Color octet leptons}

Color octet leptons ($l_{8}$) are predicted in preonic models with colored preons [31]. There are various phenomenological studies on $l_{8}$ at TeV energy scale colliders [32-39]. Resonant production of  color octet electron ($e_{8}$) and muon ($\mu_{8}$) at the FCC based lp colliders have been considered in [40] and [41] respectively. Performing similar analyses for SppC based lp colliders we obtain  mass discovedynamicry limits for $e_{8}$ and $\mu_{8}$ in $\Lambda = M_{l_{8}}$ case (where $\Lambda$ is compositeness scale) which are presented in Figs 2 and 3, respectively. Discovery mass limit value for LHC and SppC are obtained by rescaling ATLAS/CMS second generation LQ results [42, 43] using the method developed by G. Salam and A. Weiler [44]. For lepton colliders, it is obvious that discovery mass limit for pair production of $l_{8}$ are approximately half of CM energies. It is seen that $l_{8}$ search potential of SppC based lp colliders overwhelmingly exceeds that of LHC and lepton colliders. Moreover lp colliders will give an opportunity to determine compositeness scale (for details see [40, 41]).

It should be noted that FCC/SppC based lp colliders has great potential for search of a lot of BSM phenomena, such as excited leptons (see [45] for ${\mu}^*$), contact interactions, R-parity violating SUSY etc. 
\clearpage

\begin{figure}[!h]
\centering
\includegraphics[scale=0.30]{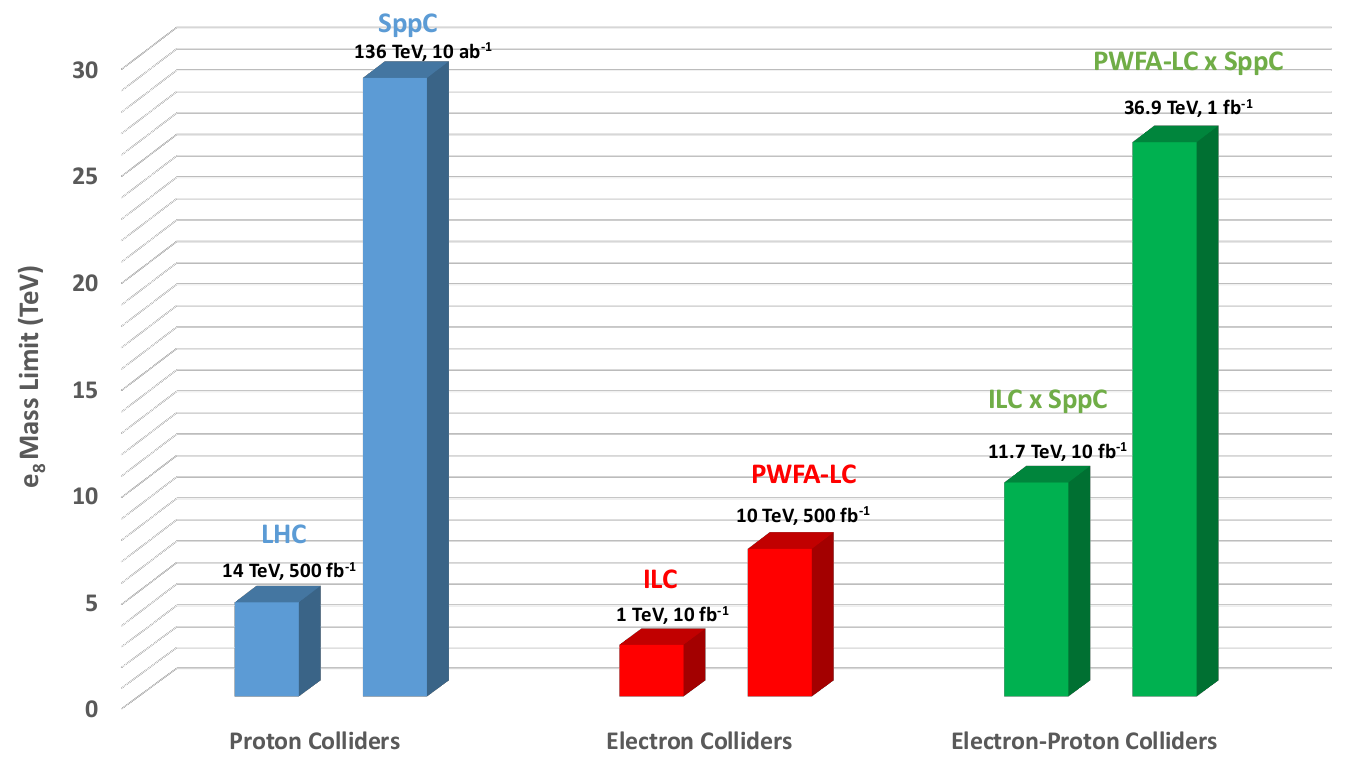}
\caption{Discovery mass limits for color octet electron at different pp, $e^+$$e^-$ and ep colliders.}
\end{figure}
\begin{figure}[!h]
\centering
\includegraphics[scale=0.30]{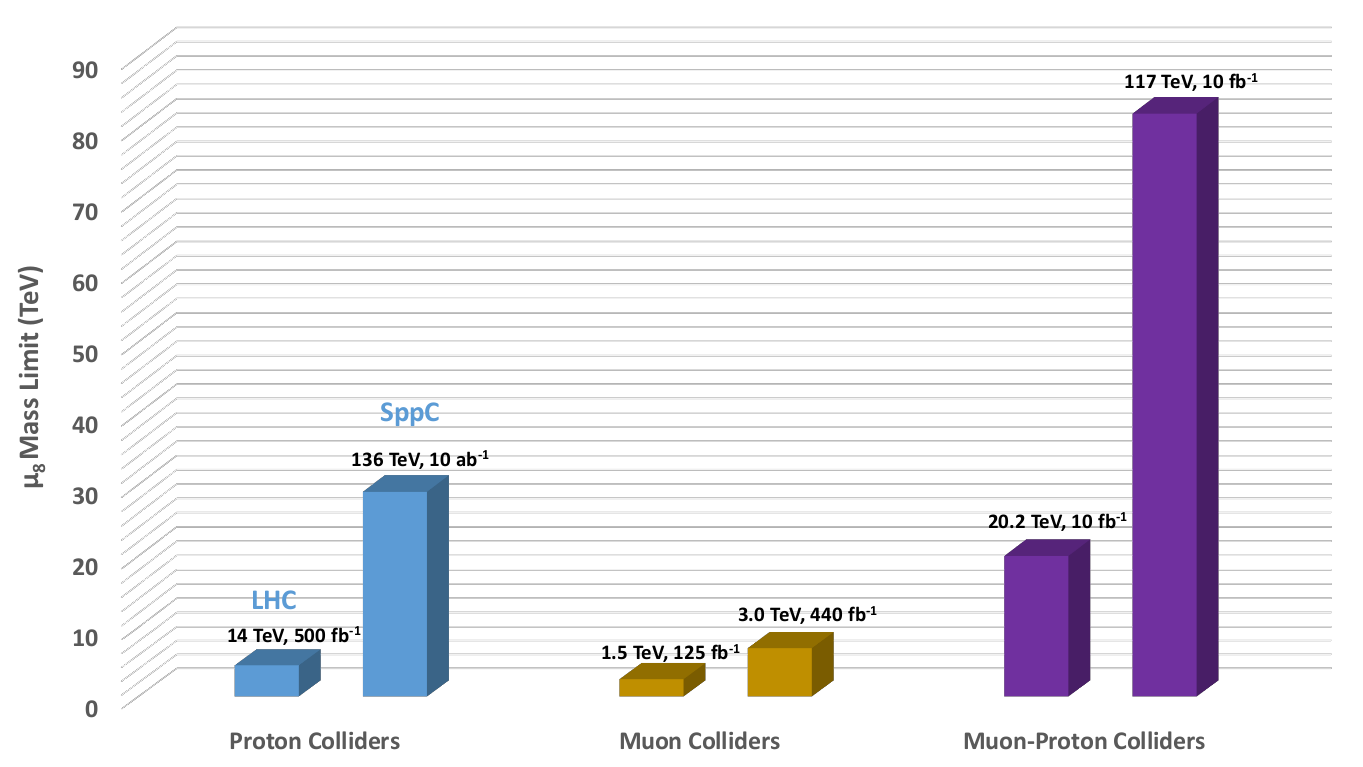}
\caption{Discovery mass limits for color octet muon at different pp, ${\mu}^+$${\mu}^-$ and ${\mu}$p colliders.}
\end{figure}

\section{Conclusion}
It is shown that construction of linear $e^{+}e^{-}$colliders (or dedicated linac) and muon colliders (or dedicated muon ring) tangential to the SppC will give opportunity to handle lepton-proton collisions with multi-TeV CM energies and sufficiently high luminosities. Concerning SM physics, these machines will certainly shed light on QCD basics. BSM search potential of lp colliders essentially exceeds that of corresponding lepton colliders. Also these type of colliders exceed the search potential of the SppC itself for a lot of BSM phenomena.

Acceleration of ion beams at the SppC will give opportunity to provide multi-TeV center of mass energy in eA and $\mu$A collisions. In addition, electron beam can be converted to high energy photon beam using Compton backdynamic-scattering of laser photons which will give opportunity to construct LC$\bigotimes$SppC based $\gamma$p and $\gamma$A colliders. Studies on these topics are ongoing. 

In conclusion, systematic study of accelerator, detector and physics search potential issues of the SppC based ep, eA, $\gamma$p, $\gamma$A, $\mu$p and $\mu$A colliders are essential to foreseen the future of particle physics. Certainly, realization of these machines depend on the future results from the LHC as well as FCC and/or SppC.   
 
\vspace{10pt}

\section*{Acknowledgments}
\addcontentsline{toc}{section}{Acknowledgement}
This study is supported by TUBITAK under the grant no 114F337.

\end{document}